\documentclass[preprint,12pt]{elsarticle}
\usepackage{hyperref}
\hypersetup{
hidelinks=true,
colorlinks=true,
linkcolor=blue,
filecolor=blue,      
urlcolor=blue,}
\usepackage{booktabs}
\usepackage{multirow}
\usepackage{threeparttable}
\usepackage{caption}

\usepackage{amssymb}
\usepackage{amsmath}

\journal{Computerized Medical Imaging and Graphics}

\begin{document}

\begin{frontmatter}



\title{RSTAR4D: Rotational Streak Artifact Reduction in 4D CBCT Using Separable 4D Convolutions}


\author[label1]{Ziheng Deng} 
\author[label2]{Hua Chen} 
\author[label3]{Yongzheng Zhou} 
\author[label1]{Haibo Hu} 
\author[label2]{Zhiyong Xu} 
\author[label4,label5]{Tianling Lyu} 
\author[label4]{Yan Xi} 
\author[label5]{Yang Chen} 
\author[label3]{Jiayuan Sun*} 
\author[label1]{Jun Zhao*} 

\affiliation[label1]{organization={School of Biomedical Engineering, Shanghai Jiao Tong University},
            city={Shanghai},
            country={China}}

\affiliation[label2]{organization={Department of Radiation Oncology, Shanghai Chest Hospital, Shanghai Jiao Tong University School of Medicine},
            city={Shanghai},
            country={China}}

\affiliation[label3]{organization={Department of Respiratory Endoscopy, Department of Respiratory and Critical Care Medicine, Shanghai Chest Hospital, Shanghai Jiao Tong University School of Medicine},
            city={Shanghai},
            country={China}}

\affiliation[label4]{organization={Jiangsu First-Imaging Tech.},
            city={Shanghai},
            country={China}}

\affiliation[label5]{organization={Research Center of Augmented Intelligence, Zhejiang Lab},
            city={Hangzhou},
            country={China}}

\affiliation[label6]{organization={School of Computer Science and Engineering, Southeast University},
            city={Nanjing},
            country={China}}

\cortext[cor1]{Corresponding authors}

\begin{abstract}

Four-dimensional cone-beam computed tomography (4D CBCT) provides respiration-resolved images and can be used for image-guided radiation therapy. However, the ability to reveal respiratory motion comes at the cost of image artifacts. As raw projection data are sorted into multiple respiratory phases, the cone-beam projections become much sparser and the reconstructed 4D CBCT images are covered by severe streak artifacts. Although several deep learning-based methods have been proposed to address this issue, most algorithms formulate it as a 2D image enhancement task, neglecting the dynamic nature of 4D CBCT images. In this paper, we first identify the origin and appearance of streak artifacts in 4D CBCT images. We find that streak artifacts exhibit a unique “rotational motion” along with the patient’s respiration, distinguishable from diaphragm-driven respiratory motion \textbf{in 4D space}. Therefore, we propose a novel \underline{4D} neural network model, \textbf{RSTAR4D-Net}, designed to address \underline{R}otational \underline{ST}reak \underline{A}rtifact \underline{R}eduction by \textbf{exploring the dynamic prior of 4D CBCT images}. Specifically, we overcome the \textbf{computational and training difficulties} of a 4D neural network. The specially designed model decomposes the 4D convolutions into multiple lower-dimensional operations and thus can efficiently \textbf{excavate the spatiotemporal information from a whole 4D image}. Additionally, a Tetris training strategy pertinent to the separable 4D convolutions is proposed to effectively train the model \textbf{using limited 4D training samples}. Extensive experiments substantiate the effectiveness of our proposed method, and the RSTAR4D-Net shows superior performance compared to other methods. The source code and dynamic demos are available at \url{https://github.com/ivy9092111111/RSTAR}.

\end{abstract}




\end{frontmatter}



\section{Introduction}
\label{sec1}
Lung cancer is one of the malignant tumors with the highest morbidity and mortality rates worldwide, posing a serious threat to human health \cite{r0}. Radiation therapy is a standard treatment for lung cancer. Modern radiation therapy equipment integrates cone-beam computed tomography (CBCT) imagers into the system (also known as on-board CBCT) to perform image-guided radiation therapy (IGRT). The in-room imaging capability of on-board CBCT improves lung tumor localization, enabling more precise radiation delivery \cite{r0.5}. However, compared with conventional CT scans, on-board CBCT scans take much longer data acquisition times (e.g., 1 minute), making them more sensitive to organ motion. A significant challenge in lung IGRT is posed by respiratory motion, which leads to blurry CBCT images and inaccurate radiation delivery.

To address respiratory motion artifacts, the 4D CBCT technique has been introduced \cite{r1,r2}. This approach synchronizes CBCT data acquisition with breathing signals. The raw projection data are then sorted into multiple phases (\textit{e.g.}, 10 phases) according to these breathing signals. By independently reconstructing each single-phase image, a multi-phase image series is generated to achieve dynamic imaging. However, the adoption of 4D CBCT involves a tradeoff between temporal resolution and spatial sampling frequency. Within each respiratory phase, the projection data exhibit an extremely sparse and non-uniform distribution \cite{r21}. Severe streak artifacts dominate the reconstructed images and hinder the clinical utility of 4D CBCT.

To enhance the quality of 4D CBCT images, two main strategies have been investigated in recent years. The first uses motion compensation methods, which model the respiratory motion pattern and subsequently integrate all projection data. The respiratory motion model can be derived from uncorrected 4D CBCT images \cite{r4}, pretreatment 4D CT images \cite{r5}, or raw projection data \cite{r6, r6.5}. However, the efficacy of the motion compensation method heavily depends on the accuracy of the estimated motion model, and the optimization of this model can be challenging. The second strategy implements compressed sensing (CS) based algorithms to reconstruct images using only phase-specific projection data. Notably, temporal correlations within 4D CBCT images have been explored, leading to the proposal of various hand-crafted regularization terms \cite{r7,r8}. However, these methods are sensitive to hyperparameters, which should be tuned on a case-by-case basis. Several hybrid methods that combine these two strategies have also been proposed \cite{r10,r11}. Nevertheless, the complexity of these algorithms may increase image reconstruction time. 

Instead of explicitly modeling the inverse process of CT imaging, recent advancements in deep learning research demonstrate that CT image reconstruction can be effectively addressed in a data-driven manner \cite{r12}. In 4D CBCT imaging, 2D convolutional neural networks (CNNs) are the most popular tool. In 2019, Jiang \textit{et~al.} proposed a symmetric residual CNN (SR-CNN) to enhance the results of CS-based algorithms \cite{r13}. Yang \textit{et~al.} adopted an adversarial learning strategy to train a group of CNNs in a phase-specific manner \cite{r14}. The idea of combining the average image (reconstructed from projection data of all phases) and single-phase image has gained widespread acceptance. A dual-encoder structure was proposed to extract relatively static and dynamic features separately to enhance the quality of 4D CBCT images \cite{r15}. Similarly, Zhi \textit{et~al.} proposed a CycN-Net with a dual-encoder structure and explored the similarity among images at adjacent phases \cite{r16}. Hu \textit{et~al.} proposed a hybrid framework named PRIOR to improve reconstructed images through an iterative algorithm \cite{r17}. These 2D methods split the 4D image into 2D inputs, considering only a small fraction of the whole 4D image during the image recovery process. The 2D outputs are then stacked along both the temporal and spatial (mostly the z-axis) dimensions to achieve the final result.

Despite the impressive abilities of data-driven algorithms in image recovery, existing 2D methods may not fully exploit the intrinsic structural priors of 4D CBCT images. Specifically, the unique imaging protocol of 4D CBCT generates a distinct dynamic pattern of streak artifacts (will be discussed in \ref{sec2}). The temporal features could be valuable for the network model to differentiate artifacts from desired anatomical structures. Additionally, respiratory motion, mainly driven by the moving diaphragm, causes notable image deformation along the z-axis. \textbf{2D methods neglect the inter-phase and inter-plane correlations} among 4D CBCT images and focus solely on in-plane image enhancement, which makes it difficult to identify artifacts and may introduce inconsistencies in the recovered image. 

In this paper, we first investigate the origin and characteristics of streak artifacts in 4D CBCT images. We find that phase-specific streak artifacts manifest in a distinctive rotational pattern within a breathing cycle. Intuitively, we show that the motion pattern of the streak artifacts differs from that of respiratory motion. This dissimilarity suggests the possibility of \textbf{decoupling them in 4d space}. Building upon this insight, we propose a \underline{\textbf{4D}} neural network model named \textbf{RSTAR4D-Net} for \underline{\textbf{R}}otational \underline{\textbf{ST}}reak \underline{\textbf{A}}rtifact \underline{\textbf{R}}eduction in 4D CBCT. Unlike conventional approaches, the RSTAR4D-Net takes the whole 4D CBCT image as the input and can concurrently consider the spatial and temporal information. Although the idea is straightforward, we argue it is nontrivial to develop a 4D network model due to higher computational costs and the demand for large-scale 4D datasets. The main methodological innovation of this paper lies in the way we instantiate the 4D network model. Specifically, the RSTAR4D-Net implements separable 4D convolutions by dividing the convolutional operation into separate dimensions, making it efficient enough to process the whole 4D image in one pass. Additionally, the specially designed model allows us to adopt \textbf{a novel Tetris training strategy} to effectively train the 4D model with very limited 4D samples. Experimental results on both simulated and real clinical data confirm the superior performance of our method.

\begin{figure}[!t]
\centerline{\includegraphics[width=\columnwidth]{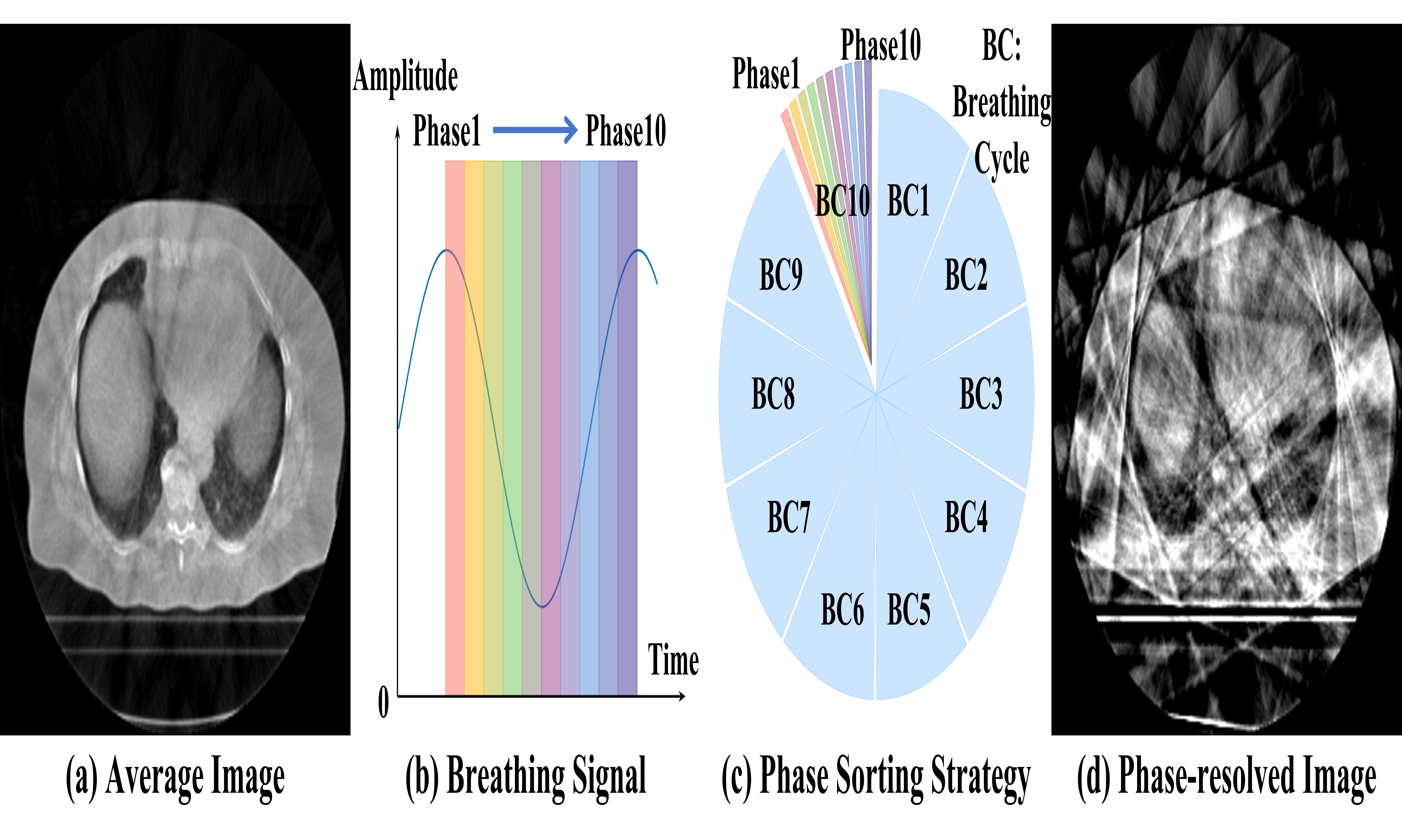}}
\caption{4D CBCT imaging with breathing signals. (a) The average image with blurry structure edges, (b) the breathing signal used to sort the projection data into different phases, (c) the phase sorting map, and (d) a phase-resolved image reconstructed from single-phase projection data, which reveals the edge of the diaphragm but suffers severe streak artifacts.}
\label{fig1}
\vspace{-0.3cm}
\end{figure}

\section{Material and methods}
\label{sec2}
In this section, we first introduce the scan protocol of 4D CBCT. We then analyze the origin and characteristics of rotational streak artifacts (RSA) in 4D CBCT images and explain the necessity of developing a 4D neural network model. Finally, we provide a detailed introduction to our RSTAR4D-Net.

\subsection{Rotational Streak Artifacts (RSA) in 4D CBCT Images}
\label{sec2-1}
In this paper, we focus on the routine 1-minute 4D CBCT scan used in IGRT. Given the typical breathing periods of patients (2-6 seconds), a 1-minute scan covers $\textit{N}_C\in[10,30]$ breathing cycles. Image reconstruction using the standard Feldkamp-Davis-Kress (FDK) algorithm with the collected projection data $\textbf{\textit{p}}$ yields a blurry average image $\textbf{\textit{f}}_{ave}$:
\begin{equation}
\textbf{\textit{f}}_{ave}=\textit{FDK}(\textbf{\textit{p}}).\label{eq1}
\end{equation}

Specifically, the regions of the moving lung and diaphragm are blurred due to respiratory motion (see Fig. 1 (a)). To capture the respiratory motion process, the breathing signal is simultaneously recorded (see Fig. 1 (b)). The projection data within each breathing cycle are then sorted into \textit{N} different respiratory phases (see Fig. 1 (c)), with each phase corresponding to a different stage of the respiration process.To reconstruct a phase-resolved CBCT image $\textbf{\textit{f}}_i$ at specific phase $\textit{i}\in[1,2,…,N]$, only the phase-specific projection data $\textbf{\textit{p}}_i$ are used to implement the FDK algorithm:
\begin{equation}\label{eq2}
\begin{split}
\textbf{\textit{f}}_{i}=\textit{FDK}(\textbf{\textit{p}}_i),\\
\textbf{\textit{f}}_{ave}={1/N}\sum_{i=1}^{N} \textbf{\textit{f}}_{i}.
\end{split}
\end{equation}

Compared to the average image, the phase-resolved image excludes projection data collected from other respiration phases, revealing clearer edges of the lung and diaphragm. However, the phase-specific projection data $\textbf{\textit{p}}_i$ exhibit troublesome sampling patterns. They have a limited sampling number and are bunched into several sparsely distributed clusters \cite{r21}. Consequently, the reconstructed CBCT image suffers from severe streak artifacts due to insufficient sampling (see Fig. 1 (d)). \textbf{Our goal is to reduce streak artifacts and restore the degraded phase-resolved images}.

\begin{figure}[!t]
\centerline{\includegraphics[width=\columnwidth]{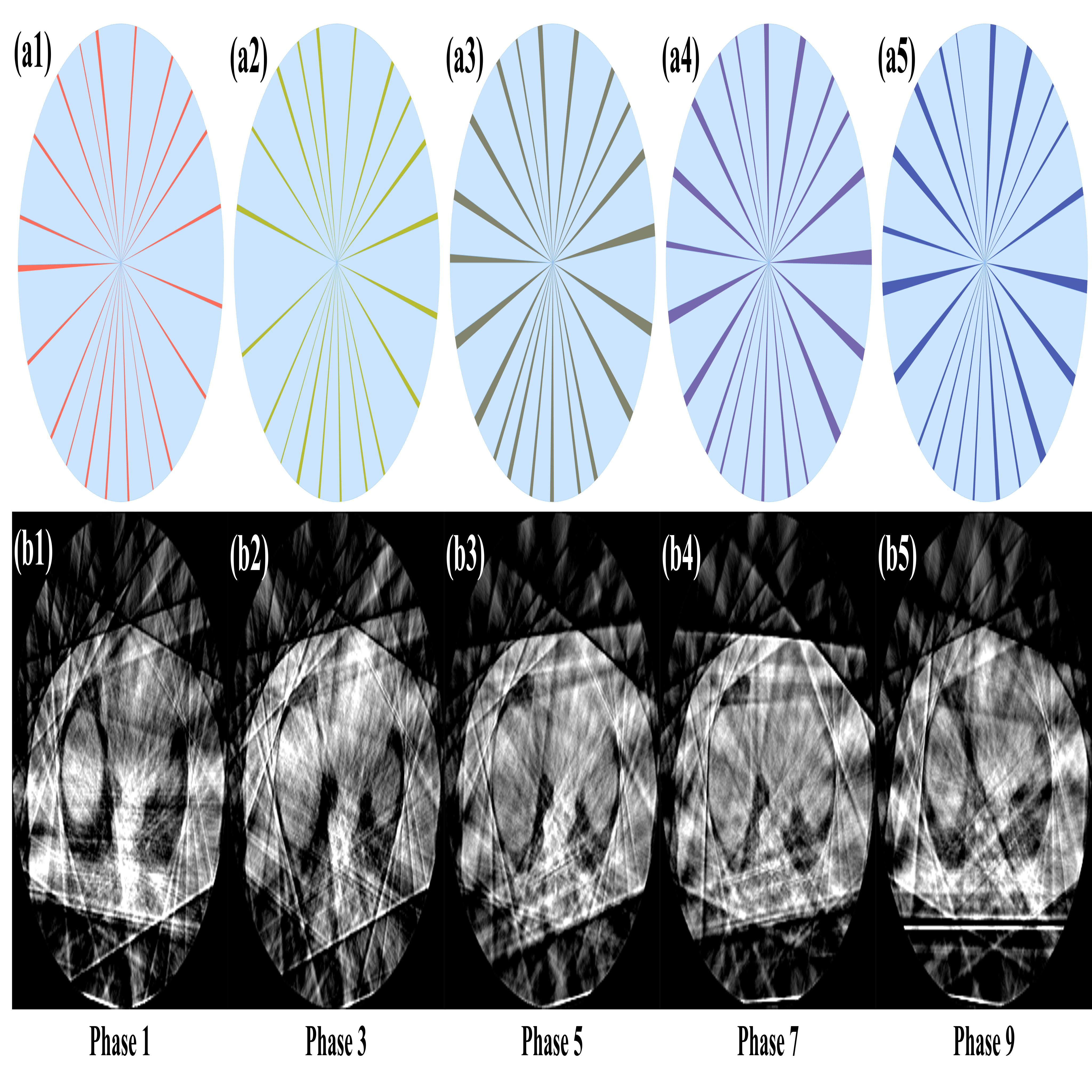}}
\caption{Illustration of rotational streak artifacts (RSA). (a1)-(a5) The projection sampling patterns at five respiratory phases, (b1)-(b5) phase-resolved images with RSA.}
\label{fig2}
\vspace{-0.4cm}
\end{figure}

Here, we analyze the distribution of streak artifacts. Streak artifacts generally arise from discontinuities in projection data \cite{r22}. In phase-specific projection data, the edges between sampled and unsampled views introduce discontinuities along the view-sampling direction, resulting in streak artifacts on each single-phase image. \textbf{The distribution of streak artifacts is closely related to the projection sampling pattern}. Without loss of generality, this phenomenon can be explained by the central slice theorem for 3D parallel-beam CT \cite{r32}. Specifically, when sparsely sampled projection data are used for image reconstruction, only a star-shaped area in the Fourier space is recovered. According to the properties of the Fourier transform, striped structures in the frequency domain thus induce streak artifacts along vertical directions in the spatial domain. 

An example is shown in Fig. 2, where a real clinical breathing signal is used to simulate a 4D CBCT scan. The first row displays the projection sampling patterns of five respiratory phases. The distributions of projection discontinuities vary from phase to phase, leading to streak artifacts with varying directions in the reconstructed images (Fig. 2 (b1)-(b5)). In general, prominent streaks span along the direction where the sampling discontinues, which indicates that the distributions of streak artifacts are correlated with projection sampling patterns.

Furthermore, given the quasi-periodic and sequential nature of respiration motion, the projection sampling pattern also evolves systematically within a breathing cycle. In essence, it appears to rotate regularly (see Fig. 2 (a1)-(a5)). As a result, the corresponding streak artifacts also rotate within the breathing cycle (see Fig. 2 (b1)-(b5)), and please also refer to our \href {https://github.com/ivy9092111111/RSTAR} {website} for video demos). We here describe the unique image artifacts in 4D CBCT as \textbf{rotational streak artifacts (RSA)}, which provides additional prior knowledge that is useful for artifact reduction. While the distribution of streak artifacts also depends on the scanned anatomical structures and may not be explicitly described with a mathematical model, we intuitively illustrate the idea through an experiment.

In Fig. 3, we compute the optical flow fields \cite{r34} among the 4D CBCT image to track the motion trajectories of densely sampled feature points \cite{r23}. Fig. 3 (a) depicts the respiratory motion in an ideal 4D CBCT image without streak artifacts. As the diaphragm primarily moves up and down (see Fig. 3 (e)), the organ motion in the axial section is minimal and mainly manifests as mild expansion and contraction. The motion trajectories are also evaluated on an average image affected by RSA (Fig. 3 (b)). The average image contains no respiratory motion, and the observed “motion” in the axial section is caused by the rotation of streak artifacts. A real 4D CBCT image exhibits a hybrid motion incorporating both respiratory motion and streak artifact motion, as shown in Fig. 3 (c). The visualization results are consistent with our analysis of RSA. Additionally, the motion trajectories of densely sampled feature points allow us to further characterize the motion patterns in dynamic images \cite{r23}. We apply the t-distributed stochastic neighborhood embedding (t-SNE) algorithm \cite{r24} to reduce the dimensionality of each motion trajectory and project it onto a two-dimensional space (Fig. 3 (d)). The result indicates the potential to \textbf{distinguish the motion patterns of anatomical structures and RSA in the spatiotemporal space}. In Fig. 3 (h), we present the average optical flow components along the x-, y-, and z-axes for each motion pattern. The respiratory motion has a dominant component along the z-axis, while RSA induces image deformation mainly in the x-y plane. This observation further underscores the significance of \textbf{considering both spatial and temporal information in 4D CBCT image recovery}.

\begin{figure}[!t]
\centerline{\includegraphics[width=\columnwidth]{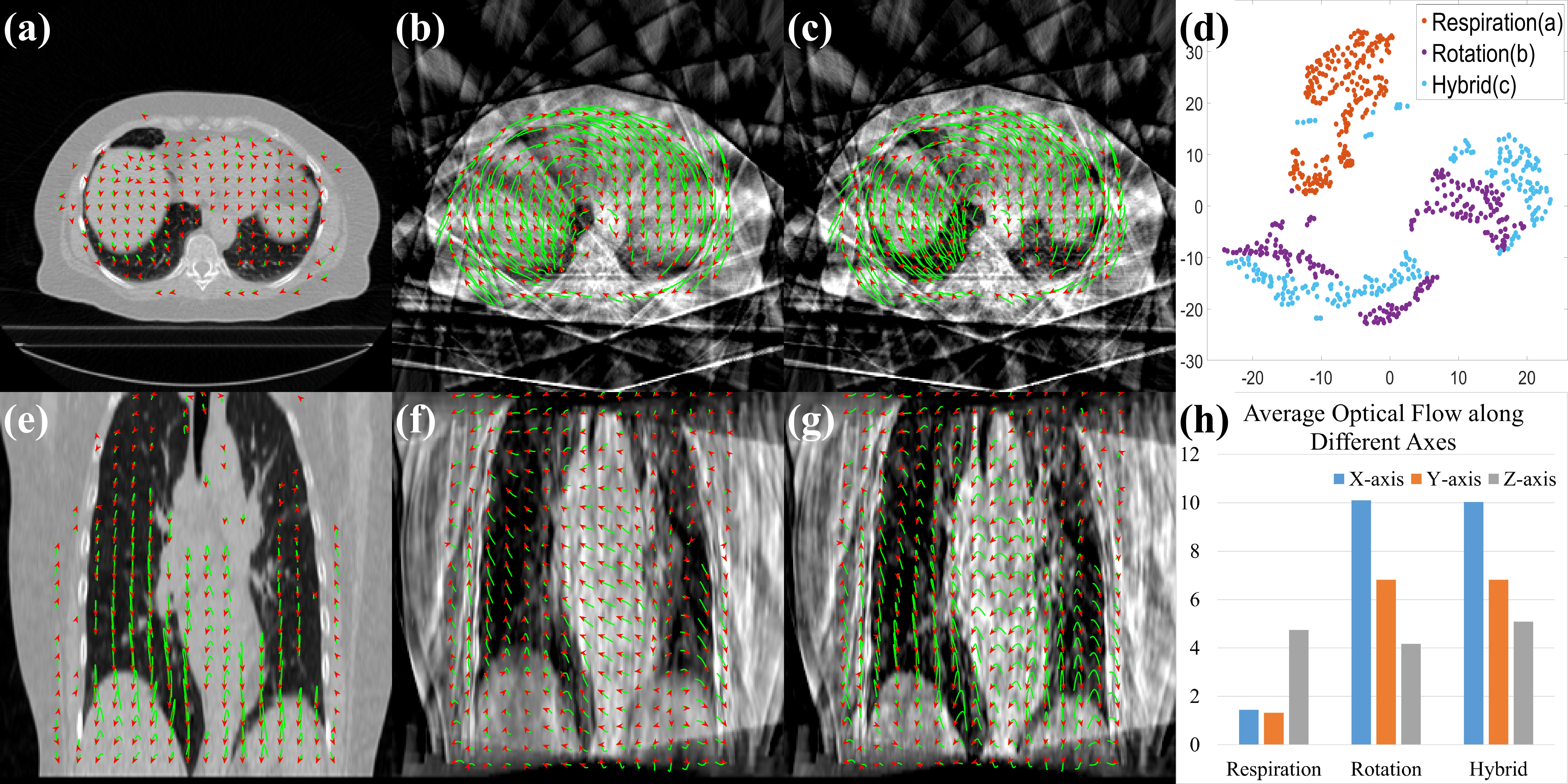}}
\caption{Motion trajectories of densely sampled feature points projected to axial and coronal planes. (a) Respiration-induced motion in the axial plane, (b) RSA-induced motion in the axial plane, (c) Hybrid motion in the axial plane, (d) t-SNE plot of the three types of motion trajectory; (e) Respiration-induced motion in the coronal plane, (f) RSA-induced motion in the coronal plane, (g) Hybrid motion in the coronal plane, (h) The average optical flow component along different axes.}
\label{fig3}
\vspace{-0.3cm}
\end{figure}

\begin{figure}[!t]
\centerline{\includegraphics[width=\columnwidth]{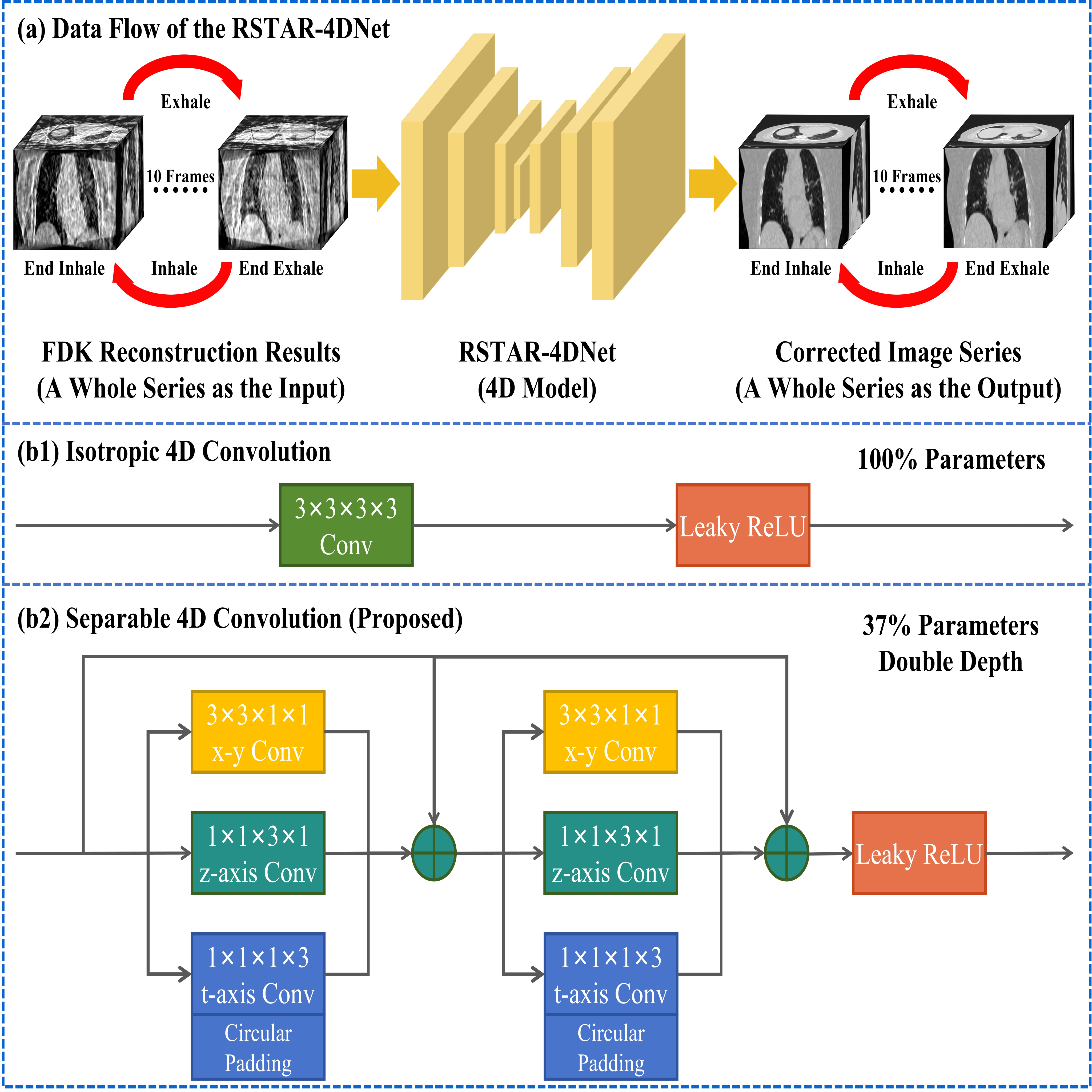}}
\caption{The design of the RSTAR4D-Net. (a) The RSTAR4D-Net takes a whole 4D CBCT image as the input. (b1-b2) 4D convolutional modules: (b1) Vanilla isotropic 4D convolutions, (b2) lightweight separable 4D convolutions (proposed).}
\label{fig4}
\vspace{-0.3cm}
\end{figure}

\subsection{RSTAR4D-Net}
\label{sec2-2}
The aforementioned discovery can serve as valuable prior knowledge and propel the 4D CBCT research. In this study, we focus on developing data-driven approaches for streak artifact reduction in 4D CBCT and formulate it as an image recovery task. \textbf{We explore an unconventional way by using a fully 4D neural network model that processes the whole 4D CBCT image directly}. Specifically, we propose the RSTAR4D-Net to recover the 4D CBCT image from the FDK reconstruction results (see Fig. 4 (a)). RSTAR4D-Net is a pure CNN with separable convolutions tailored to respiratory motion, avoiding complex designs with high system complexity. This section details our approach to effectively train a 4D model with \textbf{limited computational resources and 4D training samples}.

\subsubsection{Model Architecture}
We leverage the well-known U-Net \cite{r25} as the backbone model. Initially, we replace the original 2D convolutional filters with 4D convolutional filters to accommodate the input data (see Fig. 4 (b1)). However, using isotropic 4D convolutions is computationally expensive. For example, a 3$\times$3$\times$3$\times$3 spatiotemporal convolutional filter requires nine times the parameters and computational costs compared to a 3$\times$3 filter. Instead, we decompose the 4D convolution into multiple lower-dimensional convolutions. Specifically, our implementation decomposes the 4D convolution into a 2D spatial convolution in the x-y plane, a 1D spatial convolution in the z dimension, and a 1D temporal convolution in the temporal dimension (see Fig. 4 (b2)). This approach increases the number of parameters and computational cost by only two-thirds while achieving a 4D convolution. A similar idea has been explored to modify 3D convolutions in action recognition research for natural videos \cite{r26} and recently in large video generation models \cite{r27.5}. However, to our knowledge, this is the first attempt to develop a 4D neural network model for 4D CBCT research. Additionally, separable convolutions alleviate the training difficulty, which will be discussed in the next subsection. With the separable convolution strategy, the RSTAR4D-Net can \textbf{process a whole 4D CBCT image of size 256$\times$256$\times$90$\times$10 (height, width, slice, frame) in one pass} at the inference stage with a single consumer-grade GPU (an Nvidia RTX 3090 with 24GB memory). Furthermore, instead of using standard zeros padding, we employ circular padding for the temporal convolution to better align with the quasi-periodic nature of respiratory motion. As a result, the proposed RSTAR4D-Net demonstrates a high capacity for recurrently extracting spatiotemporal information from the 4D CBCT image.

\subsubsection{Tetris training strategy}

The paradigm of using 2D neural network models in medical imaging research is driven by the scarcity of medical images. By cropping volumetric medical images into multiple slices, a 2D dataset can be created for deep learning, even if it involves only dozens of patients. However, in this study, we aim to train a 4D neural network model capable of processing spatiotemporal features of a 4D CBCT image. Consequently, we require 4D training samples, and training deep neural networks with only dozens of samples will easily lead to the overfitting problem. Fortunately, the unique architecture of the RSTAR4D-Net allows us to implement an effective way to \textbf{train the model using only dozens of 4D CBCT images}, which we term the Tetris training strategy. As shown in Fig. 5, the proposed Tetris training strategy is carried out in two stages.

\begin{figure}[!t]
\centerline{\includegraphics[width=\columnwidth]{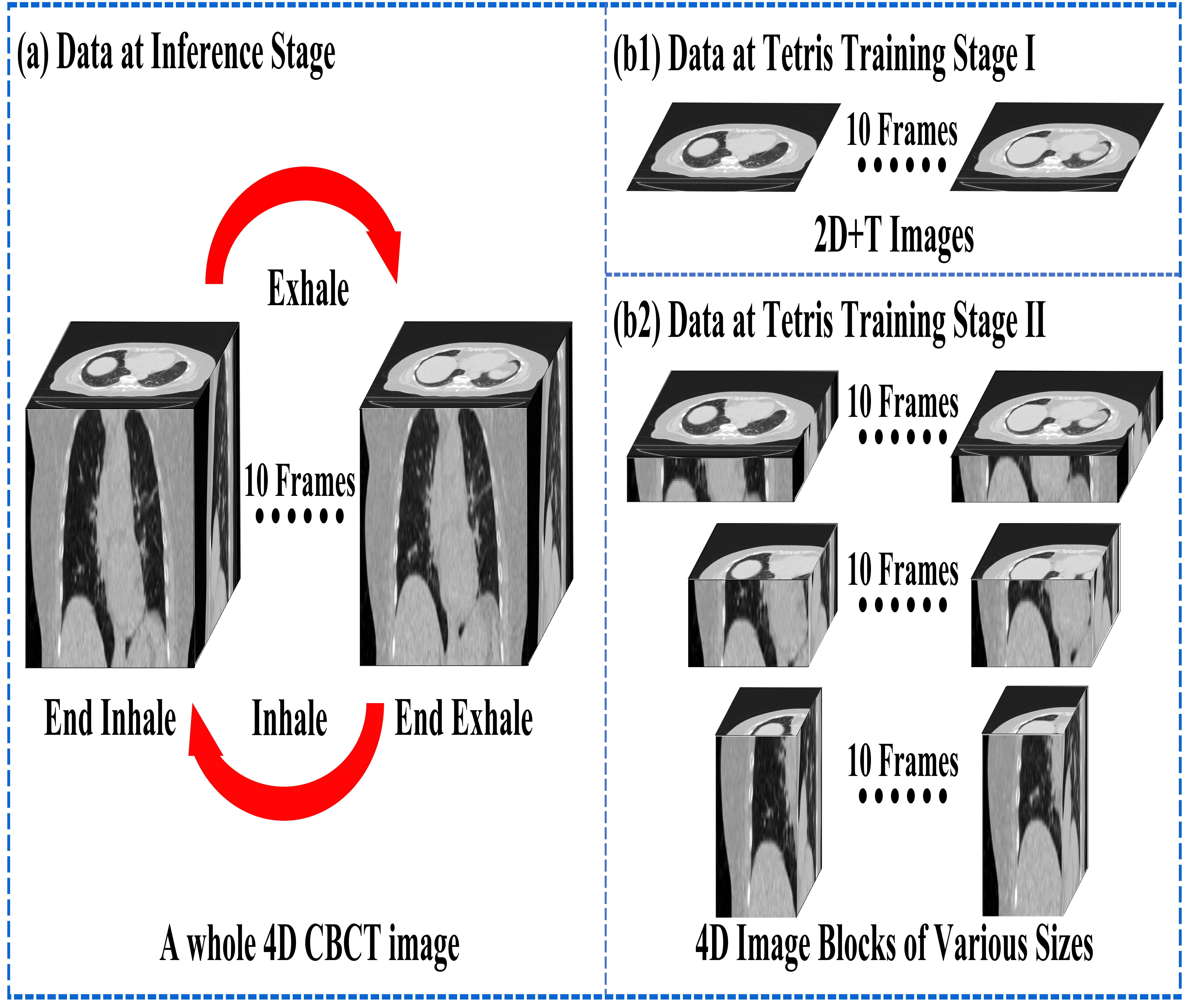}}
\caption{The Tetris training strategy. (a) The RSTAR4D-Net processes an entire 4D CBCT image at the inference stage. (b1) 2D+T images used in Tetris training stage \uppercase\expandafter{\romannumeral1}, (b2) 4D image blocks of various sizes used in Tetris training stage \uppercase\expandafter{\romannumeral2}.}
\label{fig5}
\vspace{-0.3cm}
\end{figure}

In Tetris stage \uppercase\expandafter{\romannumeral1}, we train the RSTAR4D-Net using 2D+T data. Specifically, we reformat the 4D CBCT image slice-by-slice (in axial view) to significantly increase the number of samples in the dataset. The separable convolutions in the RSTAR4D-Net include three relatively independent data flows to encode the image features along different dimensions. Thus, we can simply freeze and skip the 1D convolution module along the z-axis in Tetris stage \uppercase\expandafter{\romannumeral1} while maintaining the overall architecture of the model. As RSA primarily affects the axial plane and exhibits a unique motion pattern along the temporal dimension (as analyzed in \ref{sec2-1}), Tetris stage \uppercase\expandafter{\romannumeral1} enables the model to recognize and reduce the RSA effectively. Additionally, training on 2D+T data largely enhances data throughput and accelerates convergence.

In Tetris stage \uppercase\expandafter{\romannumeral2}, we finetune the RSTAR4D-Net using 4D image blocks of various sizes. The 4D CBCT images used in this study have a size of 256$\times$256$\times$90$\times$10 (height, width, slice, frame). In our implementation, we empirically crop these 4D images into blocks of three different sizes: 256$\times$256$\times$16$\times$10, 144$\times$144$\times$32$\times$10, and 128$\times$128$\times$64$\times$10. During this stage, we unfreeze the 1D z-axis convolution module and optimize the entire 4D model on 4D samples. As the diaphragm-driven respiratory motion leads to image deformation mainly along the z-axis (as analyzed in \ref{sec2-1}), Tetris stage \uppercase\expandafter{\romannumeral2} allows the model to accommodate respiratory motion and enhance volumetric consistency. We avoid training directly on the entire 4D CBCT image (size of 256$\times$256$\times$90$\times$10) due to high memory requirements. Additionally, the cropping step serves as data augmentation, helping to mitigate the overfitting problem.

In summary, the separable convolutions in the RSTAR4D-Net provide the flexibility to train the 4D model progressively. The Tetris training strategy enables the model to be trained on multiple image blocks of various sizes and finally process an entire 4D CBCT image. This approach makes it feasible to develop a 4D neural network model with limited computational resources and 4D training samples, which we believe will advance future research in various imaging modalities.

\subsection{Experiments}
\subsubsection{Simulated training and validation datasets}
45 high-quality 4D CT images were retrospectively collected from Shanghai Chest Hospital using a SOMATOM Definition AS CT scanner (Siemens Healthcare). Each 4D CT image includes 10 3D volumes representing different respiratory phases to cover a whole respiratory cycle. We then simulated 4D CBCT images from the 4D CT images. We used 4D CT images instead of 4D CBCT images due to their superior image quality and larger field of view. Breathing signals for 20 real clinical 4D CBCT scans were obtained from the SPARE challenge dataset \cite{r20}. Subsequently, we simulated axial CBCT scans using the 4D CT images and the breathing signals. The scan configuration was derived from the scan protocol of an On-Board imager\textsuperscript{TM} from Varian Medical System. The distances from the source and detector to the rotation center were 1000 mm and 500 mm, respectively. The detector had a physical size of 397 mm $\times$ 298 mm with 1024$\times$768 channels. The half-fan scan mode was simulated by horizontally shifting the detector to increase the field of view. The view sampling rate was 680 projection views per turn, and it took 1 minute for a 360-degree rotation. At each projection view, the specific respiratory phase was determined by the breathing signal, and a cone-bean projection was simulated from the corresponding CT volume. We performed gated FDK reconstruction to obtain 4D CBCT images. For each patient, we selected five different breathing signals to simulate five 4D CBCT scans. For each simulated scan, the reconstructed image contained 10 256$\times$256$\times$90 volumes at different phases. In total, 200 and 25 4D samples were allocated to the training and validation datasets, respectively. The 4D samples were further cropped into various shapes for neural network training, as detailed in Table 1.

\begin{table}
\label{table:1}
\tiny
\centering
\tabcolsep=0.08cm
\caption{Simulated training and validation datasets used in this study}
\begin{tabular}{cccccc}
\toprule[1.5pt]
\multirow{2}{*}{Dataset}  & \multirow{2}{*}{No. of patients} & \multirow{2}{*}{No. of 4D samples} & \multirow{2}{*}{No. of 2D samples} & No. of 2D+T samples & No. of 4D blocks \\
\multirow{2}{*}            & \multirow{2}{*}      & \multirow{2}{*}   & \multirow{2}{*}    & (Tetris Stage \uppercase\expandafter{\romannumeral1})   & (Tetris Stage \uppercase\expandafter{\romannumeral2}) \\
\midrule
Training dataset & 40 & 200 & 180000 & 18000 & 3000\\
Validation dataset & 5 & 25 & 22500 & 2250 & 375\\
\bottomrule
\end{tabular}
\end{table}

\subsubsection{Simulated test dataset}
11 high-quality 4D CT images were collected from the 4D-Lung Cancer Imaging Archive \cite{r28}. Note that the DIR-LAB dataset includes multiple scans per patient, only the earliest scan was involved in this study. Also, scans from another 9 patients were excluded due to relatively low image quality (primarily because of the poor volumetric consistency). Simulated 4D CBCT scans were generated using the same configurations as described above, each paired with an unused breathing signal. The original 4D CT images served as ground truth for evaluating the performance of different algorithms.

\subsubsection{Real clinical test dataset}
12 real clinical CBCT scans were collected from Shanghai Chest Hospital. All patients were diagnosed with thoracic tumors and required radiotherapy. Raw projection data were acquired using an On-Board imager\textsuperscript{TM} (Varian Medical System) under a 1-minute half-fan scan protocol. Although breathing signals were not available, we estimated them directly from the projection data using the Amsterdam Shroud technique \cite{r33}. The projection data at each view were sorted into specific respiratory phases based on the estimated breathing signals, with an amplitude-based sorting strategy to partially relieve the irregular breathing problem. We then reconstructed phase-resolved 4D CBCT images from the raw projection data using the gated FDK algorithm. After that, the reconstructed images were processed with the above methods, and the results were used for qualitative evaluation. Due to the lack of ground-truth images, the quantitative evaluation was performed in the projection data domain.

\subsubsection{Comparison Methods}
Five algorithms were involved for comparison, including 1) Gated FDK method, 2) PICCS method \cite{r7}, 3) MSD-GAN \cite{r14}, 4) CycN-Net \cite{r16}, and 5) Prior-Net \cite{r17}. The gated FDK method is widely used in clinical practice, which also serves as the input to all the deep neural networks. The PICCS method is a compressed sensing-based method and takes the average image as a prior image. The MSD-GAN method trains 10 phase-specific deep neural networks to reduce the streak artifacts at each respiratory phase. Both the CycN-Net and the Prior-Net are deep neural networks employing a dual-encoder architecture to fuse information from the average and phase-resolved images. For a fair comparison, we also concatenated the average image with the degraded image to serve as the input to both the MSD-GAN model and the proposed RSTAR4D-Net. While all three deep learning-based comparison methods utilized 2D CNNs, the proposed RSTAR4D-Net considers the whole 4D CBCT image, addressing correlations among different slices and respiratory phases. Note that in the original paper of the MSD-GAN and the Prior-Net, the neural network was further combined with either a motion compensation strategy or an iterative CT reconstruction algorithm. To avoid confounding the performance evaluation of neural network models with additional strategies, we omitted these extra modules from our comparative studies. Nevertheless, the RSTAR4D-Net can be seamlessly integrated into various frameworks.

\subsubsection{Implementation Details}
We trained the RSTAR4D-Net, CycN-Net, and Prior-Net using the pixelwise L1 loss. The MSD-GAN was trained with the multiscale-discriminator GAN loss combined with the pixelwise L1 loss according to the original paper. We used the PyTorch toolbox to implement all the deep learning-based methods. The training was performed on an Nvidia RTX 3090 GPU with 24 GB memory. Adam algorithm was used to optimize the network models, with a learning rate initially set to $10^{-4}$ and slowly decreased to $10^{-5}$ with 30 epochs. The batch size was set to 1. 

The structural similarity (SSIM) index and the root mean squared error (RMSE) were adopted to quantitatively evaluate various methods on the simulated test dataset. Additionally, we segmented the lung regions and the tumor regions as regions of interest (ROI) to better present the results. The lung regions are segmented using an in-house algorithm. The tumor regions were determined from RTSTRUCT files in the 4D-Lung Cancer Imaging Archive, where the contours of the tumors were annotated by clinicians.

\section{Results}
\subsection{Experiments on the Simulated Dataset}

\begin{table}
\scriptsize
\centering
\tabcolsep=0.08cm
\caption{Quantitative evaluation (means ± sds) of various methods}
\begin{tabular}{ccccccc}
\toprule[1.5pt]
\multirow{2}{*}{Metric} & \multirow{2}{*}{Gated-FDK} & \multirow{2}{*}{PICCS} & \multirow{2}{*}{MSD-GAN} & \multirow{2}{*}{CycN-Net} & \multirow{2}{*}{PriorNet} & RSTAR4D-Net \\
\multirow{2}{*}         & \multirow{2}{*}      & \multirow{2}{*}      & \multirow{2}{*}   & \multirow{2}{*}    & \multirow{2}{*}   & (proposed) \\
\midrule
SSIM-Tumor(\%)$\uparrow$ & 31.38±8.90 & 69.41±8.11 & 72.87±9.97 & 75.71±8.83 & 78.33±8.37 & \textbf{84.33±5.84}\\
SSIM-Lung(\%)$\uparrow$ & 24.37±6.68 & 65.73±5.20 & 69.85±6.27 & 72.37±5.74 & 75.51±5.52 & \textbf{80.74±5.12}\\
SSIM-Global(\%)$\uparrow$ & 16.46±3.22 & 69.35±3.82 & 71.26±3.30 & 70.28±3.70 & 72.73±3.26 & \textbf{75.77±2.77}\\
RMSE-Tumor(HU)$\downarrow$ & 321.43±83.03 & 128.07±26.00 & 131.11±42.43 & 120.61±37.56 & 103.50±34.14 & \textbf{75.03±21.50}\\
RMSE-Lung(HU)$\downarrow$ & 288.61±46.54 & 118.99±16.57 & 109.43±20.71 & 103.81±17.65 & 85.18±15.48 & \textbf{66.68±12.90}\\
RMSE-Global(HU)$\downarrow$ & 365.26±46.64 & 80.13±11.61 & 60.23±8.41 & 61.18±6.14 & 49.32±4.75 & \textbf{42.54±3.90}\\
\bottomrule
\end{tabular}
\vspace{-0.3cm}
\end{table}

Quantitative evaluation was conducted for all methods on the simulated dataset. We used the SSIM and RMSE indices to evaluate the reconstruction accuracy for the tumor regions, the lung regions, and the whole body. The quantitative results are shown in Table 2. Our proposed RSTAR4D-Net yields images closest to the ground truth images, with the best SSIM and RMSE results.

\begin{figure*}[!t]
\centerline{\includegraphics[width=16cm]{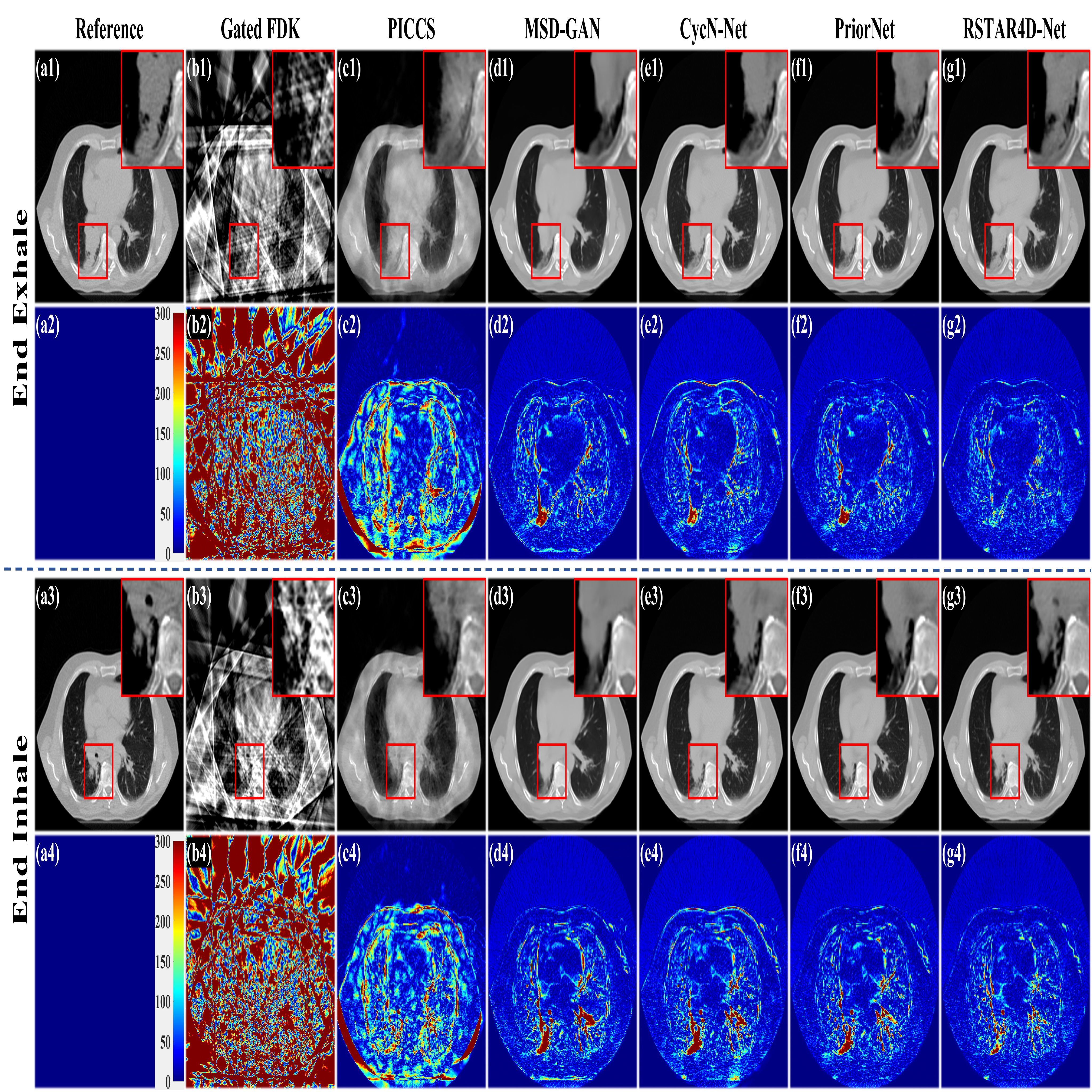}}
\caption{Axial reconstruction results of the simulated dataset arranged as follows: (a) Reference, (b) Gated FDK, (c) PICCS, (d) MSD-GAN, (e) CycN-Net, (f) PriorNet, (g) RSTAR4D-Net. The first and the third rows present the reconstructed images with zoom-in tumor regions, and the second and the fourth rows present the corresponding difference images in absolute value. The display window is [-1000, 500] HU for reconstructed images, [-450, 150] HU for zoom-in tumor regions, and [0, 300] HU for absolute difference images.}
\label{fig6}
\vspace{-0.3cm}
\end{figure*}

\begin{figure*}[!t]
\centerline{\includegraphics[width=16cm]{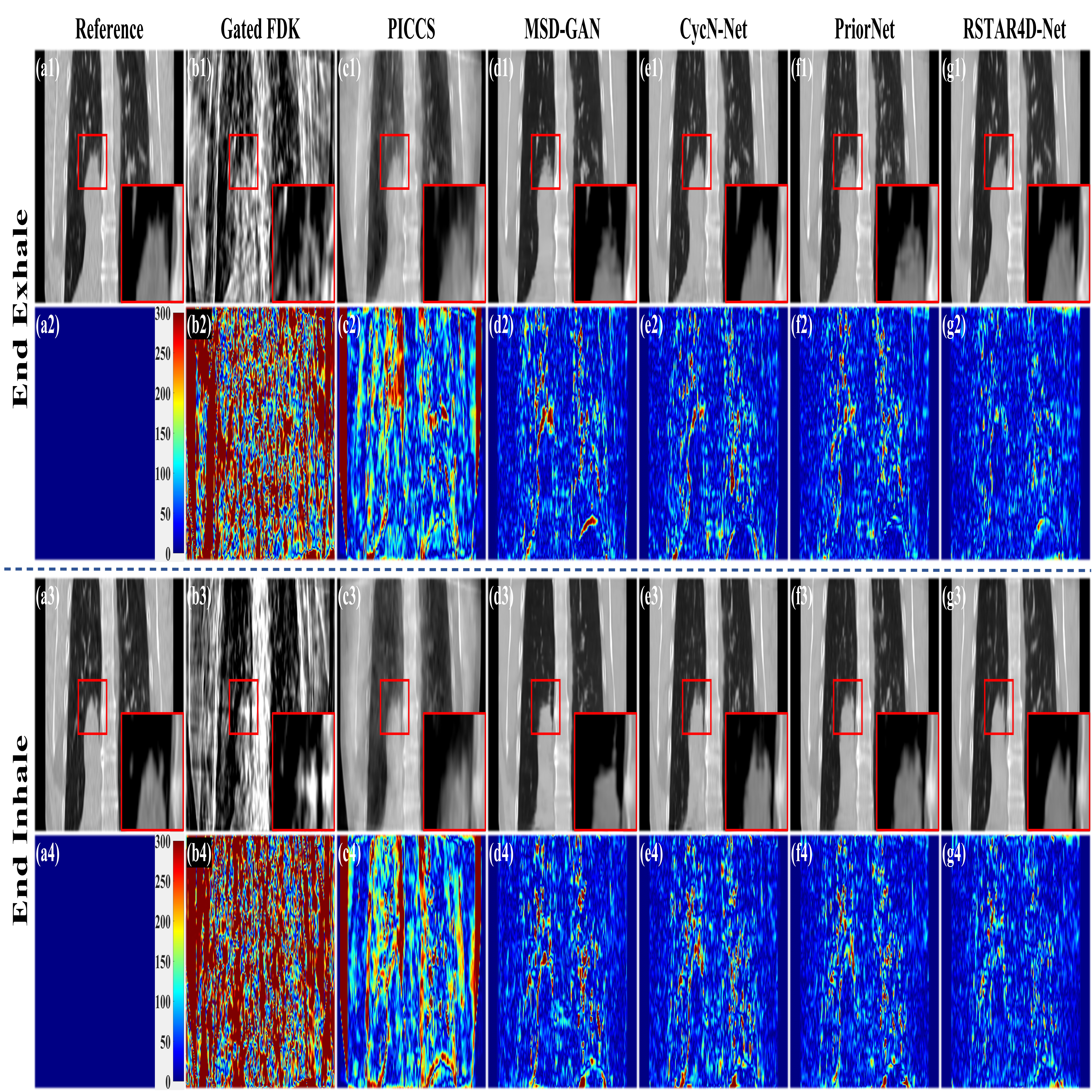}}
\caption{Coronal reconstruction results of the simulated dataset arranged as follows: (a) Reference, (b) Gated FDK, (c) PICCS, (d) MSD-GAN, (e) CycN-Net, (f) PriorNet, (g) RSTAR4D-Net. The first and the third rows present the reconstructed images with zoom-in tumor regions, and the second and the fourth rows present the corresponding difference images in absolute value. The display window is [-1000, 500] HU for reconstructed images, [-450, 150] HU for zoom-in tumor regions, and [0, 300] HU for absolute difference images.}
\label{fig7}
\vspace{-0.3cm}
\end{figure*}

Fig. 6 presents the axial views of the reconstructed images from different methods. Severe streak artifacts are observed in the gated FDK results. The PICCS method removes most streaks but tends to over-blur the images. Generally, all four deep learning-based methods provide images with increased quality. However, recovering some detailed structures, especially tumors, remains challenging. The comparison methods (MSD-GAN, CycN-Net, and PriorNet) struggle to distinguish desired anatomical structures from streak artifacts because only limited information is available. As shown in Fig. 6 (d1-f1), the tumor regions are nearly invisible under a narrower window. In contrast, the proposed RSTAR4D-Net utilizes the entire series of CBCT images and better recovers the shape of the tumor (Fig. 6 (g1)). Similarly, at the end-inhale phase, the proposed RSTAR4D-Net achieves superior reconstruction results among all methods.

Fig. 7 presents the coronal views of the reconstructed images from different methods. As the proposed RSTAR4D-Net performs image recovery directly in 4D space, the reconstruction results exhibit better volumetric consistency. The inter-slice respiratory motion is naturally resolved by our 4D neural network model, making the edges of the tumor and the diaphragm much clearer.

\begin{table}
\scriptsize
\centering
\tabcolsep=0.08cm
\caption{Quantitative results (means) from the ablation study}
\begin{tabular}{ccccccc}
\toprule[1.5pt]
2D Model & 2D+t Model & 4D Model & Tetris Stage \uppercase\expandafter{\romannumeral1} & Tetris Stage \uppercase\expandafter{\romannumeral2} & SSIM-Tumor(\%)$\uparrow$ & SSIM-Lung(\%)$\uparrow$\\
\midrule
$\checkmark$ &  \quad &  \quad &  \quad&  \quad & 77.48 & 74.55\\
\quad  & $\checkmark$ & \quad & $\checkmark$  & \quad  & 80.91 & 77.73\\
\quad  & \quad  & $\checkmark$ & \quad  &  \quad & 68.20 & 67.02\\
\quad  & \quad  & $\checkmark$ & \quad  & $\checkmark$ & 82.20 & 78.60\\
\quad & \quad & $\checkmark$ & $\checkmark$ & $\checkmark$ & \textbf{84.33} & \textbf{80.74}\\
\bottomrule
\end{tabular}
\end{table}

We further conducted an ablation study on the simulated dataset to demonstrate the effectiveness of the proposed training strategy. The quantitative results of the ablation study are shown in Table 3. The 2D+T model significantly outperforms the 2D model, underscoring the importance of incorporating dynamic image series. However, a 4D model simply trained on a limited-scale 4D dataset yields unsatisfactory results. The proposed Tetris training strategy is essential for better training the 4D model, resulting in a clear performance gain.

In our study, we also found that it is recommended to pretrain the model on 2D+T data instead of 2D data because the temporal information is quite important for the recovery of 4D CBCT image. Moreover, further finetuning the model on the whole 4D CBCT images brought little performance gain. However, the proposed method can be used in other 4D scenarios, and these attempts may find specific roles in different tasks.

\begin{table}
\scriptsize
\centering
\tabcolsep=0.08cm
\caption{The number of parameters and computational cost of different network models}
\begin{tabular}{ccccc}
\toprule[1.5pt]
\multirow{2}{*}{Metric} & \multirow{2}{*}{MSD-GAN} & \multirow{2}{*}{CycN-Net} & \multirow{2}{*}{PriorNet} & RSTAR4D-Net \\
\multirow{2}{*}       & \multirow{2}{*}   & \multirow{2}{*}    & \multirow{2}{*}   & (proposed) \\
\midrule
No. of parameters (million) & 113.77 & 3.55 & 13.69 & 2.35\\
FLOPs (G/slice) & 1120.30 & 43.57 & 734.68 & 118.93\\
Test time (ms/slice)& 22.99 & 13.02 & 42.08 & 13.95\\
\bottomrule
\end{tabular}
\vspace{-0.3cm}
\end{table}

The number of parameters and computational costs of all four neural network models are listed in Table 4. Thanks to the separable convolution strategy, the RSTAR4D-Net can implement 4D convolution at an affordable computational cost. Compared to the MSD-GAN and PriorNet, the proposed method is both more effective and efficient. 

\subsection{Experiments on the Real Clinical Dataset}

\begin{figure*}[!t]
\centerline{\includegraphics[width=16cm]{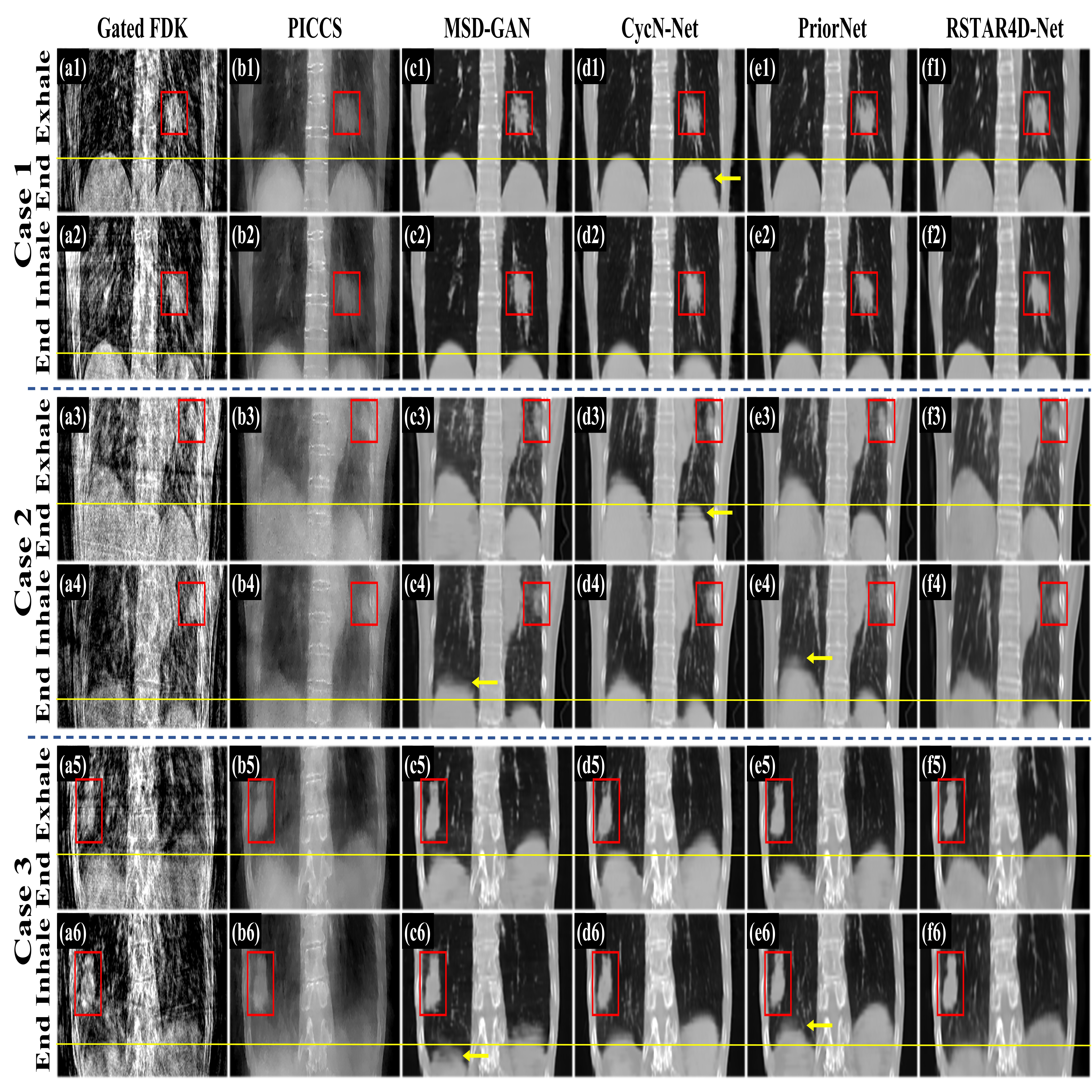}}
\caption{Reconstruction results of the real clinical dataset (cases 1-3) in coronal views arranged as follows: (a) Gated FDK, (b) PICCS, (c) MSD-GAN, (d) CycN-Net, (e) PriorNet, (f) RSTAR4D-Net. The display window is [-1000, 500] HU for reconstructed images. The red boxes indicate the tumors. The horizontal yellow lines indicate the diaphragms. The yellow arrows denote the volumetric inconsistencies along the z-axis.}
\label{fig8}
\vspace{-0.3cm}
\end{figure*}

In this section, we present the results of applying our RSTAR4D-Net to real clinical datasets, showcasing its performance in a practical setting. In Fig. 8, we present the reconstructed results of three real clinical CBCT scans (cases 1, 2, and 3). Although ground truth images are unavailable, we evaluate the performance of each method by analyzing the positions of the tumor and diaphragm \cite{r37}. The gated FDK algorithm reconstructs phase-resolved images using only the projection data collected at the specific phase, resulting in severe streak artifacts due to insufficient sampling. However, the degraded images still help locate the position of the moving tumor and diaphragm (please see Fig.8 (a1)-(a6)), as inter-phase interference is omitted. Thus, we take the gated FDK images as a reference to determine if the methods correctly removed streak artifacts while preserving anatomical structures. For each sub-figure, the tumor is denoted with a red box, and the relative position of each red box remains the same. Horizontal yellow lines mark the diaphragm’s position. While all deep learning-based methods reduce streak artifacts on real clinical 4D CBCT images, Fig. 8 reveals that anatomical structures may be inaccurately estimated by the comparison methods. Specifically, volumetric inconsistency is observed in the diaphragm regions (noted by the yellow arrows). Additionally, a lower diaphragm’s position is observed at the end-exhale phase (please see Fig. 8 (d3), (e3), (c5)), while a higher diaphragm’s position is observed at the end-inhale phase (please see Fig. 8 (e2), (d6), (e6)). These results suggest that the comparison methods may overly rely on the average image, failing to estimate the exact amplitude of respiratory motion. In contrast, the proposed RSTAR4D-Net successfully recovers the positions of the moving organs.

\begin{figure*}[!t]
\centerline{\includegraphics[width=16cm]{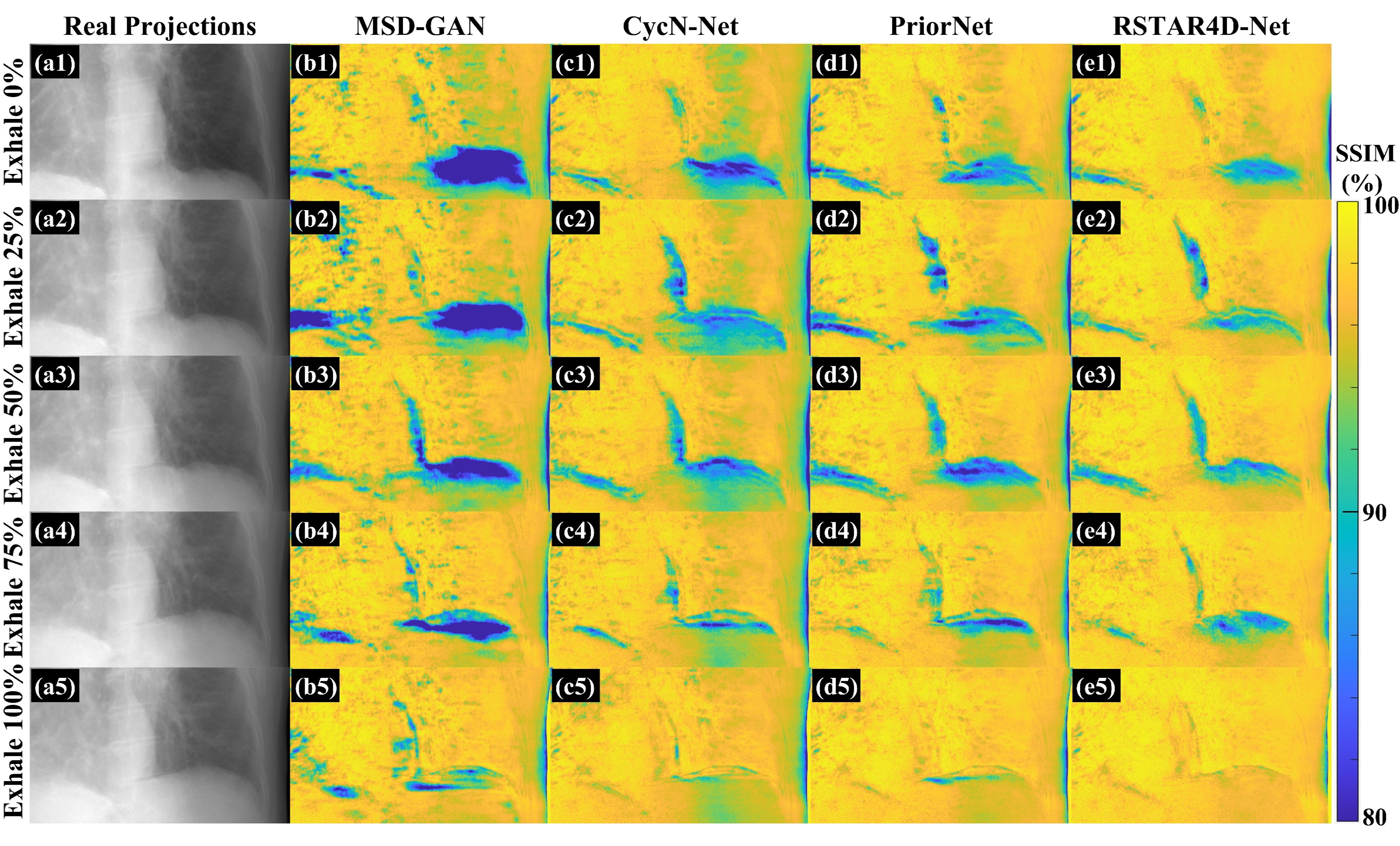}}
\caption{Comparison between the real and recovered projection data (clinical case 4). (a) Real Projection data. (b-e) SSIM maps of different methods: (b) MSD-GAN, (c) CycN-Net, (d) PriorNet, (e) RSTAR4D-Net.}
\label{fig9}
\vspace{-0.3cm}
\end{figure*}

To further evaluate the deep learning-based reconstructions, we conducted a quantitative analysis in the projection domain. We simulated CBCT scans based on the breathing signals to reproject the reconstructed results into the projection domain, enabling a comparative analysis between real and recovered projection data. We use the SSIM and the Normalized Cross Correlation Coefficient (NCC) to characterize the similarity of the projection data. Twelve scans were included in the real clinical dataset and the results are shown in Table 5. 
The visualization of the comparison is also shown in Fig. 9, where we present the entire exhalation process of clinical case 4. The first column displays the real projections at five different view angles, which correspond to different respiratory phases. The SSIM maps are displayed in the next four columns, each corresponding to a deep learning method. The results demonstrate the superiority of RSTAR4D-Net in maintaining the consistency of projection data across different respiratory phases.

\begin{table}[!t]
\scriptsize
\centering
\tabcolsep=0.08cm
\caption{Quantitative results (means) on the real clinical dataset}
\begin{tabular}{ccccc}
\toprule[1.5pt]
\multirow{2}{*}{Metric} & \multirow{2}{*}{MSD-GAN} & \multirow{2}{*}{CycN-Net} & \multirow{2}{*}{PriorNet} & RSTAR4D-Net \\
\multirow{2}{*}       & \multirow{2}{*}   & \multirow{2}{*}    & \multirow{2}{*}   & (proposed) \\
\midrule
SSIM(\%)$\uparrow$ & 95.60 & 95.86 & 96.65 & \textbf{97.02}\\
NCC(\%)$\uparrow$ & 97.88 & 97.85 & 98.30 & \textbf{98.75}\\
\bottomrule
\end{tabular}
\vspace{-0.3cm}
\end{table}

\section{Discussion and Conclusion}
In this paper, we proposed a novel deep learning-based method for reducing rotation streak artifacts (RSA) in 4D CBCT. Our approach integrates domain knowledge into a data-driven framework. Initially, we investigated the origin and appearance of streak artifacts in 4D CBCT images. Thereafter, we identified the unique motion pattern of RSA, motivating us to decouple the artifacts from the anatomical structures in the spatiotemporal domain. Subsequently, we introduced the RSTAR4D-Net to recover the 4D CBCT images. The RSTAR4D-Net leverages separable 4D convolution and a temporally circular padding strategy to adapt to the physiological nature of respiratory motion. Extensive experiments validate the effectiveness of the proposed method. The proposed RSTAR4D-Net is a lightweight and flexible model, we posit that incorporating advanced modules (e.g., temporal transformer \cite{r35}, spatiotemporal implicit neural representation \cite{r36}) may further enhance the performance. Furthermore, we proposed the Tetris training strategy to effectively train a 4D neural network model with only a few 4D training samples. We believe the proposed method could be generalized to various 4D imaging techniques, such as dynamic MR, PET, and SPECT imaging.

With our method, during a 1-minute scan, clear 4D CBCT images can be obtained without the need for breath holding. The 4D CBCT images enable physicians to better understand the respiratory motion of the lesions and facilitate precise lesion localization during radiation therapy. The proposed method may bring along many new application scenarios. For example, the image-guided percutaneous lung biopsy is performed using CBCT. To acquire clear CBCT image, general anesthesia is needed for better respiratory control. With the proposed method, it may be possible for free-breathing CBCT scan and the procedures can be performed under moderate sedation with local anesthesia.

While our experiments demonstrate the efficacy of RSTAR4D-Net in reducing streak artifacts on real clinical CBCT images, it is essential to acknowledge that clinical data often follow a long-tailed distribution. Generalizing our method to more complex cases requires further investigation. The study of stabilizing deep tomographic reconstruction \cite{r30} has inspired us to combine the neural network model with traditional CT reconstruction techniques to enhance robustness and interpretability. Given the generic and efficient architecture of RSTAR4D-Net, it can serve as a plug-and-play module to benefit traditional methods. We plan to further extend the RSTAR4D-Net in our future studies.

\section*{Acknowledgments}
This work was supported in part by the Major Research Plan of the National Natural Science Foundation of China (No.92059206), the National Key R\&D Program of China (2022YFC2408500), the Shanghai Jiao Tong University Medical Engineering Cross Research Funds (YG2024ZD27, YG2022QN059), and the Key R\&D Program in Jiangsu Province of China (BE2021703, BE2022768).




\bibliographystyle{elsarticle-num-names.bst}
\biboptions{numbers,sort&compress}
\bibliography{reference}

\end{document}